\documentclass[aps,prx,amsfonts,amsmath,amssymb,raggedbottom,longbibliography,reprint,superscriptaddress,citeautoscript]{revtex4-2}

\usepackage{graphicx} 
\usepackage{float}
\usepackage{color}
\usepackage{bm}
\usepackage{hyperref}
\usepackage{todonotes}
\usepackage{verbatim}
\usepackage{soul}
\usepackage{glossaries}
\usepackage{sidecap}
\usepackage{hyperref}
\usepackage{ulem}
\hypersetup{
    colorlinks=true,
    linkcolor=blue,
    filecolor=magenta,
    urlcolor=blue,
    citecolor=blue,
}

\begin{document}

\title{Landau theory of the density wave transition in trilayer Ruddlesden-Popper nickelates}

\author{M. R. Norman}
\email{norman@anl.gov}
\affiliation{Materials Science Division, Argonne National Laboratory, Lemont, IL 60439}

\date{\today}

\begin{abstract}
This paper presents a Landau treatment of the incommensurate density wave transition observed in trilayer Ruddlesden-Popper nickelates and uses this to address the nature of the transition.  The data are consistent with a spin driven transition with the distinct intertwining of charge and spin order due to being in or proximate to the first order transition region of the Landau phase diagram.  From this approach, one also obtains an understanding of the variation of the transition temperature with rare earth size, pressure, and oxygen isotope substitution.
\end{abstract}

\maketitle

\section{Introduction}
The recent discovery of high temperature superconductivity in both bilayer \cite{Sun23} and trilayer \cite{Zhu24} Ruddlesden-Popper (RP) nickelates under pressure  has renewed interest in the nature of the density wave found at ambient pressure in both materials.  This is understandable, since in cuprates, for instance, the superconductivity emerges upon doping a parent phase that is an insulating antiferromagnet, indicating a magnetic origin of the pairing \cite{AndersonRVB,NagaosaRMP,ScalapinoRMP}.  In the RP nickelates, there has been an active debate concerning the relation between the density wave state at ambient pressure and the superconducting state at high pressure, including some observations that samples which do not exhibit a density wave at ambient pressure also do not exhibit superconductivity at high pressure \cite{Shi24,Shi25}.  To understand these observations, it would be helpful to have a better understanding of the density wave state and its relation to the structure of these materials.

In this paper, the focus will be on the trilayer material as its density wave state has been studied in great detail by a combination of neutron and x-ray scattering given the existence of large single crystals \cite{Zhang20,Anjana23}. What is found is both a charge density wave (CDW) and a spin density wave (SDW) that are intertwined in the sense that both condense at the same temperature and both appear to be primary order parameters from the perspective of the Landau theory of phase transitions.  Although this is a metal to metal transition, most of the Fermi surface seems to be removed at the transition as is apparent from the large change in the Hall coefficient when going through the transition \cite{Li21,Pan22} and the nearly complete energy gap found in STM measurements \cite{Li25}.  As for the order of the transition, this has been of some debate.  The neutron and x-ray data are consistent with a continuous transition \cite{Zhang20}. Specific heat \cite{Huangfu20,ZhangPRM,Rout20} and $\mu$SR data \cite{Khasanov25a} are consistent with a first order or weakly first order transition instead, but with little evidence for hysteresis.  Some transport data, though, do indicate hysteretic behavior \cite{Pan22}.

What drives the density wave transition, spin or charge, has also been of some debate.  For context, the CDW wavevector is twice the SDW wavevector \cite{Zhang20} as is typical for a variety of materials ranging from cuprates \cite{Tranquada95} to chromium \cite{FawcettRMP}.  In band structure calculations, there is distinct evidence for nesting at the SDW wavevector, but not at the CDW one \cite{Zhang20,Jia25}.  This would imply that spin is the driver.  But the sensitivity of the transition temperature to isotope substitution of oxygen has been taken to suggest a charge origin instead \cite{Khasanov25b}.

To investigate these questions, we consider a Landau theory of the density wave transition.  This approach has been previously used to address analogous results \cite{Lee11} for the closely related nickelate perovskites (the n=$\infty$ limit of the Ruddlesden-Popper series).  It provides a useful framework for thinking about the density waves, with the analysis presented here consistent with a spin driven transition but with coupling to charge causing the transition temperature to vary with rare earth size and oxygen isotope substitution.

\section{Landau Theory}
When considering a Landau approach, the first question concerns the coupling between the spin and charge order parameters. The lowest order coupling between the two is linear in the CDW and quadratic in the SDW order parameter given the two to one relation between their ordering vectors \cite{Zhang20}.  Given this, and the absence of harmonics of the SDW and CDW wavevectors in the data \cite{Zhang20}, the Landau free energy can be written as \cite{Zachar98}:
\begin{eqnarray}
F & = & a_2M^2(q_{SDW})+a_4M^4(q_{SDW})+b_2P^2(q_{CDW}) \nonumber \\
& & +b_4P^4(q_{CDW}) -\gamma_0 M^2(q_{SDW})P(q_{CDW}) \nonumber
\end{eqnarray}
noting that as stated above, $q_{CDW}=2q_{SDW}$, where the sign of $q$ is implicit (each term of the free energy has an overall wavevector of zero modulo a reciprocal lattice vector).  We do not allow for $M$ and $P$ to be complex since the observed density waves are linear (rather than helical), and only one wavevector each is considered (modulo its sign) since the density waves are unidirectional.  Although formally $M$ is a vector ~\cite{Zachar98,Shi23}, the vector nature is not important for the arguments of this paper so we treat $M$ as a scalar.  Redefining the coefficients as $a=a_2/\sqrt{a_4}$, $b=b_2/\sqrt{b_4}$ and $\gamma=\gamma_0/(\sqrt{a_4}b^{1/4})$, and dropping the momenta which are implicit, we end up with the simple expression:
\begin{equation}
F = aM^2+M^4+bP^2+P^4-\gamma M^2P
\end{equation}
One can easily see from this equation that the coefficient $\gamma$ can also be absorbed by renormalizing the quadratic coefficients $a$ and $b$ by $1/\gamma^2$ \cite{Zachar98}. The resulting phase diagram \cite{Zachar98} is then universal and is illustrated in Figure 1.

\begin{figure}
\centering
\includegraphics[width=0.75\columnwidth]{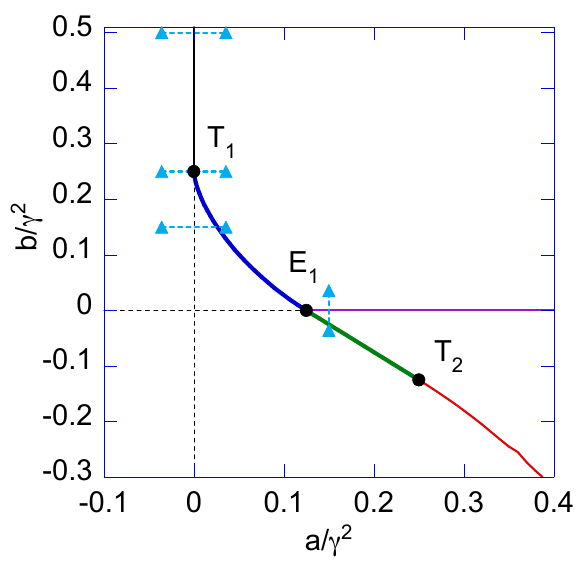}
\caption{Landau phase diagram assuming a linear-quadratic coupling term between the SDW and CDW.  Here, the $x$ axis is $a/\gamma^2$ where $a$ is the coefficient of the quadratic SDW term in the free energy, $F$, and the $y$ axis is $b/\gamma^2$ where $b$ is the coefficient of the quadratic CDW term, with $\gamma$ the coefficient of the coupling term.  The first order phase line is bounded by the points $T_1$ (0,1/4) and $T_2$ (1/4,-1/8) and is shown as a thicker curve, with the two order parameters condensing simultaneously between $T_1$ and $E_1$ (1/8,0). Above $T_1$ (with $x$=0), both order parameters also condense simultaneously but in a continuous fashion.  $E_1$ marks the beginning of a split transition.  To the right of $E_1$, at the upper transition ($y$=0) only the CDW condenses, with the SDW condensing at the lower transition (first order to the left of $T_2$, second order to the right).}
\label{fig1}
\end{figure}

An important point to realize is that the physics is independent of the sign of $\gamma$, as this can be compensated for by having a sign change of $P$ relative to $M$.  This is unlike the commonly considered biquadratic coupling term ($M^2P^2$) where a negative sign is needed to get intertwined order \cite{Holakovsky73}.  Moreover, the phase diagram for the biquadratic case is not universal since one cannot scale away the coefficient of the coupling term like one can do for the linear-quadratic case.  That is, the ratio of the coupling coefficient to the quartic coefficient is relevant, and moreover sixth order terms are also needed for stability of the theory \cite{Holakovsky73}.  In general, a negative sign is usually not found microscopically for the biquadratic term \cite{Etxebarria10}, and it also does not account for data in the perovskite nickelate case whereas the linear-quadratic term gives a good description \cite{Post18,Shi23}.  Having said that, it should be remembered that a biquadratic term can exist and potentially influence the results presented here.  Note that the layered nature of the trilayer material is not apparent in this simple Landau formalism.  The reason is that the observed SDW only occurs on the outer planes which are related by symmetry \cite{Zhang20}.  For the CDW, it appears to be have the same value in each plane \cite{Zhang20}.  Therefore, we can treat the layer indices as implicit in Equation 1.  We also ignore the rare earth moments which come into play at lower temperatures (a Landau analysis of the interplay of nickel and rare earth moments in the Pr material can be found in Ref.~\onlinecite{Anjana23}).

Taking the derivative of $F$ with respect to $M$, one finds
\begin{equation}
2M(a+2M^2-\gamma P)=0
\end{equation}
The two solutions are $M$=0 and $M^2=(\gamma P-a)/2$.  For the first, then either $P$=0 (paramagnetic phase) or $P^2=-b/2$.  The latter defines the horizontal phase line to the right of $E_1$ in Figure 1.  For the second solution, solving for $P$ from Equation 1 and inserting this expression for $M^2$, one finds
\begin{equation}
P^3+(b/2-\gamma^2/8)P+\gamma a/8=0
\end{equation}
This has six potential solutions.  Focusing on the four real ones, in practice one finds that only two of them are free energy minima.  Setting $Q=(\gamma^2/4-b)/6$ and $R=\gamma a/16$, one finds either
\begin{equation}
P=-2\sqrt{Q}\cos(\theta/3+2\pi/3), \theta=\cos^{-1}(R/\sqrt{Q^3})
\end{equation}
that defines the first-order line between $T_1$ and $T_2$ and the second-order line beyond $T_2$, and
\begin{equation}
P=-sign(R)(|R|+\sqrt{R^2-Q^3})^{1/3}
\end{equation}
that defines the (second-order) transition line above $T_1$.

\section{Results}
When considering the phase diagram, important questions are the dependence of the transition temperature on the free energy coefficients and the temperature evolution of $M$ and $P$.  Without the coupling terms, the CDW would condense for $b$=0 and the SDW for $a$=0 (dashed lines in Figure 1).  The coupling term boosts the transition temperature along the curve between $T_1$ and $E_1$ in Figure 1.  To make further progress, let us first assume that the transition is SDW driven as conjectured in Ref.~\onlinecite{Zhang20}.  Under this assumption, $b$ is independent of temperature, that is, one is taking horizontal cuts of the phase diagram in Figure 1 (diagonal cuts were considered in Ref.~\onlinecite{Post18}).
Writing $a$ as $r_{SDW}(T-T_{SDW})$ one can see that the boost in the transition temperature $T_c$ relative to $T_{SDW}$ in the region of the phase diagram between $T_1$ and $E_1$ scales as $\gamma^2/r_{SDW}$.  A plot of this quantity is shown in Figure 2 (blue curve) for $\gamma=r_{SDW}=2$ with $T_{SDW}$=140 K (the last chosen for relevance to the trilayer materials).  For this choice of parameters, one sees a $T_c$ boost as large as 35 K at the point $E_1$ in the phase diagram.  The analogous CDW driven case (vertical cuts) in the region of the phase diagram between $T_1$ and $E_1$ is shown by the red curve in Figure 2.  This would correspond to setting $b$ as $r_{CDW}(T-T_{CDW})$.

\begin{figure}
\centering
\includegraphics[width=0.75\columnwidth]{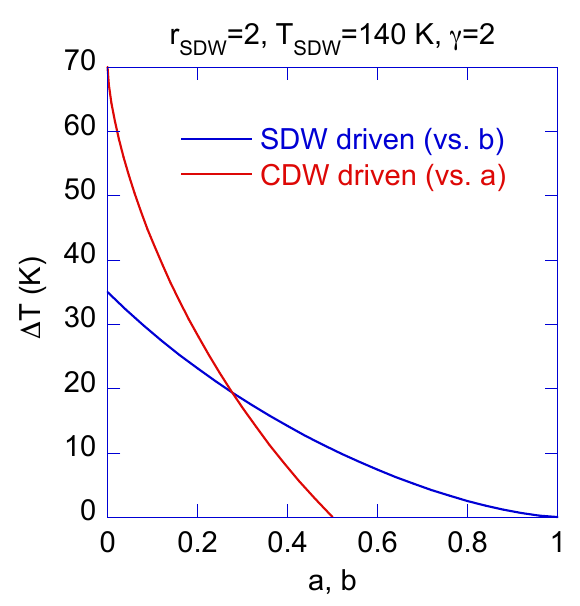}
\caption{$\Delta T = T_c - T_{SDW}$ (blue curve) as a function of $b$, the coefficient of the quadratic CDW term in $F$ for $\gamma$=2 and $r_{SDW}$=2 where $a$, the coefficient of the quadratic SDW term is $r_{SDW}(T-T_{SDW})$.  $b$=0 corresponds to the point $E_1$ in Figure 1, $b$=1 to the point $T_1$.
The red curve is $\Delta T = T_c - T_{CDW}$ as a function of $a$ for $\gamma$=2 and $r_{CDW}$=2 where $b$ is $r_{CDW}(T-T_{CDW})$. $a$=0 corresponds to the point $T_1$ in Figure 1, $a$=0.5 to the point $E_1$.}
\label{fig2}
\end{figure}

In Figure 3, we show the variation of $M$ and $P$ versus $T$ for several representative cases where either $a$ is varied (a-c) and one case where $b$ is varied (d), with these cuts of the phase diagram shown in Figure 1 as short dashed lines bounded by triangles.
In regards to the critical exponents, above the point $T_1$ in the phase diagram, sufficiently close to $T_c$, one finds that $M$ scales as $\sqrt{T_c-T}$ and $P$ as $T_c-T$.  Far enough away from $T_1$ this scaling occurs for an appreciable range in $T$ as seen in chromium \cite{FawcettRMP}, as presented in Figure 3a.  This linear in $T$ behavior for $P$ has not been observed in the trilayer materials \cite{Zhang20}. But at $T_1$, one can see from the above equations that $P$ scales as $(T_c-T)^{1/3}$ for all $T$ with $M$ scaling as $(T_c-T)^{1/6}$ very close to $T_c$ crossing over to $\sqrt{T_c-T}$ far enough away, as presented in Figure 3b.  So near $T_1$, the effective critical exponents for $M$ and $P$ are sensitive to the temperature range they are defined from.  In Figure 3c, a cut of the phase diagram is shown between $T_1$ and $E_1$ where $M$ and $P$ simultaneously condense at a first order transition, whereas in Figure 3d, a vertical cut of the phase diagram is shown between $E_1$ and $T_2$ where the CDW condenses first in a second order fashion, with the SDW condensing at a lower transition in a first order fashion (with an additional first order jump of the CDW).  Similar behavior to Figure 3d would be seen for a more general cut as long as it was not strictly horizontal.

\begin{figure}
\centering
\includegraphics[width=\columnwidth]{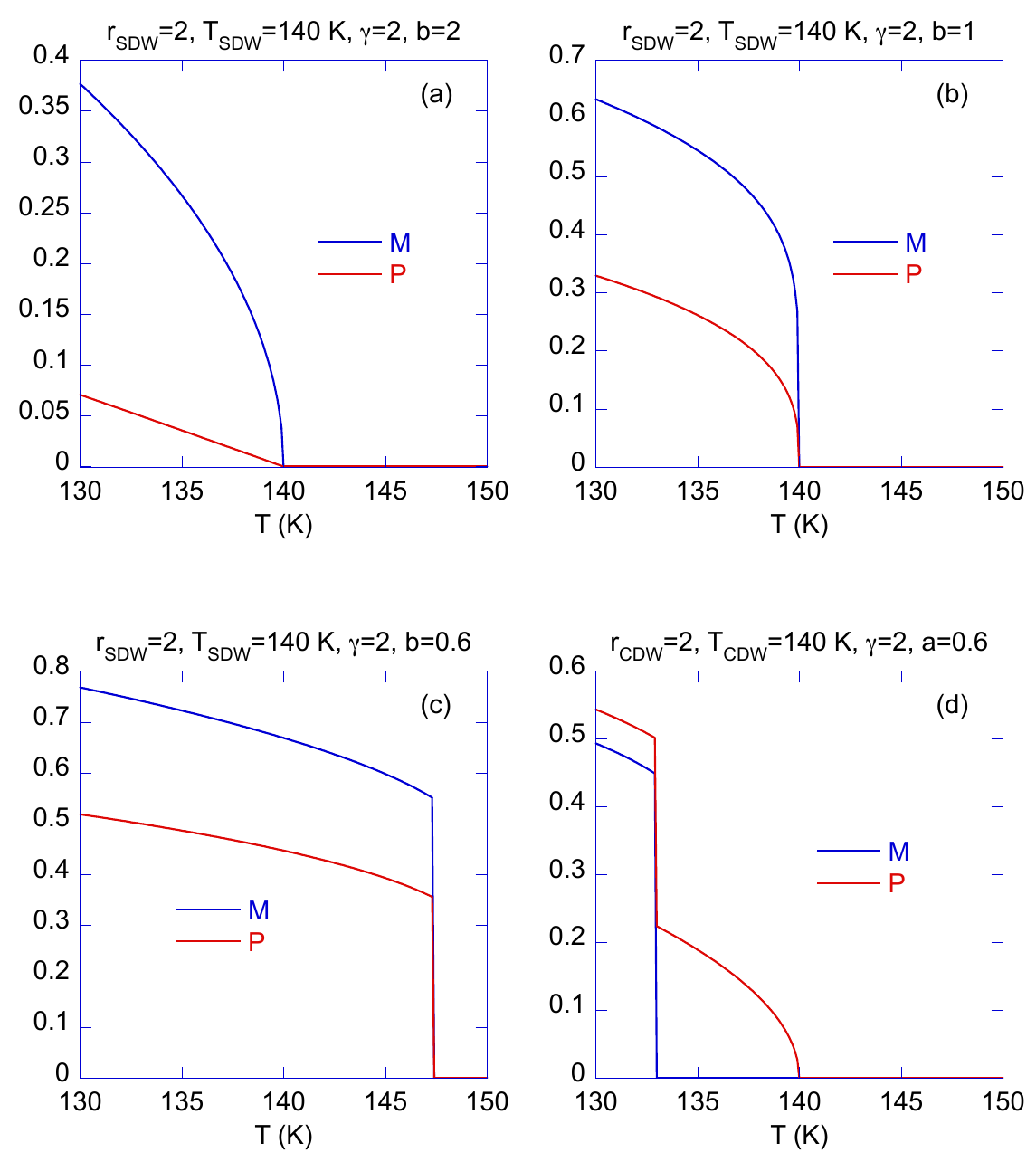}
\caption{$M$ (SDW) and $P$ (CDW) order parameters for four cuts of the phase diagram of Figure 1.  For (a)-(c), $b$ is held fixed (horizontal cuts), for (d), $a$ is held fixed (vertical cut).  (a) is above the point $T_1$ in Figure 1, (b) through the point $T_1$, (c) between $T_1$ and $E_1$, and (d) between $E_1$ and $T_2$.  These phase diagram cuts are shown in Figure 1 as short dashed lines bounded by triangles.}
\label{fig3}
\end{figure}

In the first order region, one has hysteresis (not shown in Figure 3).  The hysteretic region of the phase diagram is shown in Ref.~\onlinecite{Shen21} and plots of the order parameter versus $T$ in this region have been shown in Ref.~\onlinecite{Post18}.  In Figure 4, we show the variation of $F$ along the diagonal cut of the phase diagram ($a$=$b$) to illustrate the point that the paramagnetic solution ($M$=$P$=0) is defined by a weak free energy minimum below the instability temperature.  This means in practice that hysteretic effects should also be weak.

\begin{figure}
\centering
\includegraphics[width=0.9\columnwidth]{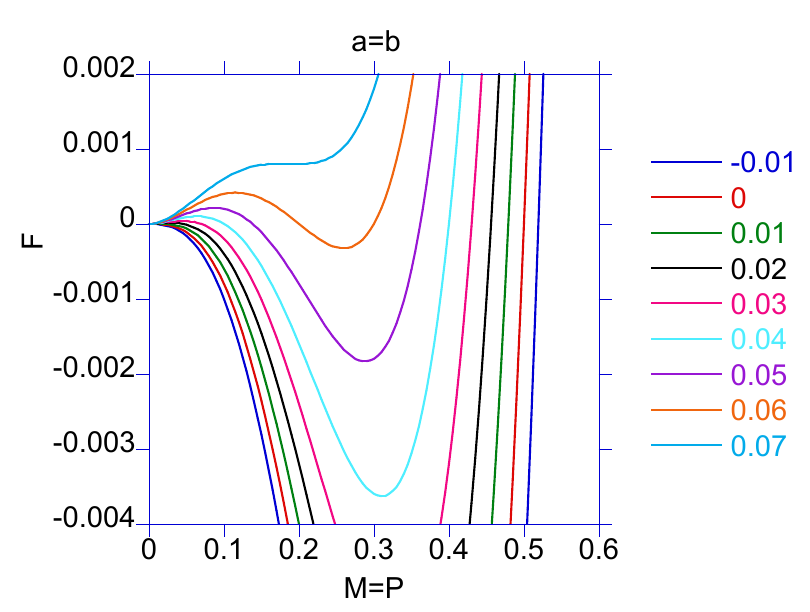}
\caption{Free energy $F$ versus the order parameter ($M$=$P$) for a diagonal cut of the phase diagram of Figure 1 that goes through the origin.  Each curve corresponds to a different value of $a$=$b$.}
\label{fig4}
\end{figure}

\section{Implications}
In nickelate perovskites \cite{Medarde98}, the metal-insulator transition has a linear relation with respect to $\cos(\phi)$ where $\phi$ is the tilt angle of the NiO$_6$ octahedra, defined as $(180-\theta)/2$ where $\theta$ is the Ni-O-Ni bond angle connecting the NiO$_2$ planes.  In Figure 5, we show an analogous plot for the trilayer materials, where the same linear relation is found for the density wave transition (a similar plot, but versus the tolerance factor, can be found in Ref.~\onlinecite{HuangfuPRR}).  Referring to Figure 2, this variation can be understood in that as the tilt angle increases, $b$ should decrease (i.e., moving towards a CDW instability), driving $T_c$ to a higher value (if one is below $T_1$ in Figure 1).  This is also consistent with the fact that from specific heat, the first-order nature of the transition in the trilayer materials seems more evident as the rare earth ion site decreases.  That is, as the tilt angle increases, $b$ decreases, and one moves further away from $T_1$ in the phase diagram and so deeper into the first-order region (as is evident from the $\Delta T$ plot in Figure 2).  In contrast, as pressure is applied, the tilt decreases and so $T_c$ also decreases, as also seen in pressure experiments \cite{Khasanov25a}.  This is also reflected in the isotope experiment, where, since $b$ should be proportional to $1/M$ where $M$ is the oxygen mass, then $T_c$ should increase upon substitution of $^{18}O$ for $^{16}O$ as is observed \cite{Khasanov25b}.  The magnitude of the isotopic $T_c$ shift ($\sim$2 K) is comparable to that seen in the perovskite nickelates \cite{Medarde98} and the same magnitude is evident from Figure 2 (since $b$ would be reduced by 11\%).  In the perovskites, a split phase transition (meaning to the right of $E_1$ in Figure 1) is seen for rare earth ions smaller than Nd \cite{CatalanoRPP} where the CDW condenses first and the SDW second.  So far, a split transition is not seen in the trilayer materials, but then again samples have not been synthesized with rare earths smaller than Nd. The observation of a split transition, if it should occur, would not be consistent with a strictly horizontal cut of the phase diagram in Figure 1.

\begin{figure}
\centering
\includegraphics[width=0.75\columnwidth]{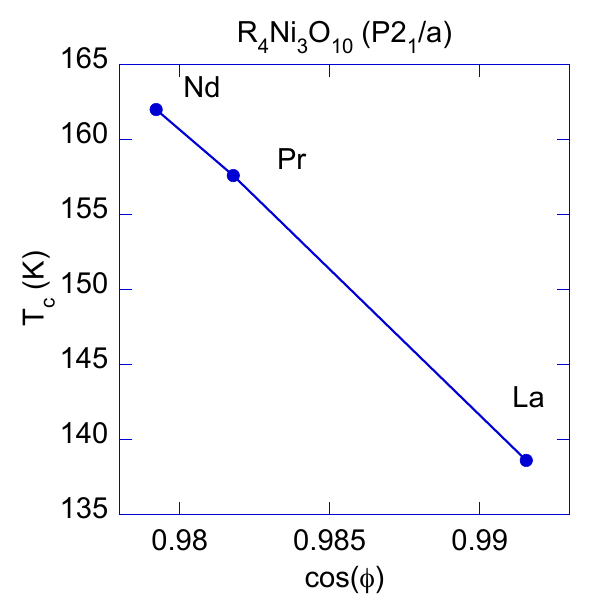}
\caption{Variation of $T_c$ versus the cosine of the octahedral tilt angle, $\phi=(180-\theta)/2$ where $\theta$ is the Ni-O-Ni bond angle between layers.  For La and Pr, the values were taken from Ref.~\onlinecite{ZhangPRM}.  For Nd, the tilt angle was taken from Ref.~\onlinecite{Olafsen00} and $T_c$ from Ref.~\onlinecite{Li21}.}
\label{fig5}
\end{figure}

On the whole, the experimental results seem largely consistent with horizontal (or near horizontal) cuts of the phase diagram in Figure 1, with the nearness to $T_1$ in Figure 1 dependent on what the order of the phase transition turns out to be.  If continuous, one should be just above $T_1$, otherwise $P$ will not appear to be a primary order parameter.  If discontinuous, as the evidence here suggests, then one is probably not too far below $T_1$.  This `horizontal cut' scenario would be in line with an SDW driven metal to metal transition, probably originating from nesting as suggested in Ref.~\onlinecite{Zhang20}, where $b$ varies as the tilt angle, oxygen isotope mass, etc., is changed.  As noted before, the density wave is sinusoidal in nature (that is, only the fundamental harmonic is seen for both the SDW and CDW), unlike stripes in cuprates where higher harmonics have been observed.  Moreover, this SDW driven scenario is also consistent with recent results that do not see any phonon anomalies at the CDW wavevector in the trilayer material \cite{Jia25}.  It will be interesting to see whether this `SDW driven' behavior persists as the rare earth size decreases.  That is, whether the parameter $b$ ever becomes negative as seems to occur in the perovskites for rare earths smaller than Nd where the transition splits into an upper metal-insulator transition (charge disproportionation) and a lower magnetic transition \cite{CatalanoRPP}, as would be the case if more general cuts than horizontal were taken in Figure 1 to the right of $E_1$.
In that context, in a recent Landau treatment of the perovskites \cite{Shi23}, the cuts used to reproduce their extensive phase diagram were nearly horizontal (with an angle of 8.5$^\circ$), in general agreement with the findings here.
 
Interestingly, in the bilayer material, it is claimed that a split transition does occur, but with the SDW condensing first \cite{Khasanov24}.  This is not consistent with Figure 1.  To obtain this type of behavior, a biquadratic coupling term must be invoked instead.

\section{Summary}

The phase diagram illustrated in Figure 1 is a convenient way of thinking about the trilayer nickelates and how their density waves vary as a function of rare earth size, pressure, and isotope substitution.  A case has been made that the current data are consistent with horizontal cuts of the phase diagram shown in Figure 1 (i.e., SDW driven), but more general cuts \cite{Post18,Shi23} of the phase diagram cannot be ruled out.  If other behavior is seen (like the SDW condensing first as has been suggested for the bilayer materials), then the picture presented in this paper has to be supplemented by a biquadratic coupling term.  Still, the important point is that even if the transition is SDW driven, the CDW order parameter can appear to be primary in the Landau sense as long as one is either in the first order region of the phase diagram (between points $T_1$ and $E_1$ in the phase diagram of Figure 1) or one is close enough to point $T_1$ in the second order region.  This clears up a major mystery that was raised in Ref.~\onlinecite{Zhang20}.  Note that above $T_1$, the phase transition temperature is independent of $b$, the coefficient of the quadratic CDW term in the free energy.  Rather, in the data, $T_c$ is seen to vary as the rare earth size is varied, pressure is varied, or upon isotope substitution, again consistent with being in the first order region bound between the points $T_1$ and $E_1$ in Figure 1 as illustrated in Figure 2.  It will be interesting to see if the rare earth size is further decreased (Sm and beyond) whether a split transition is seen where the CDW condenses first, or more generally, charge disproportion that might lead to a metal-insulator transition as occurs in the nickelate perovskites.

\begin{acknowledgments} 
This work was supported by the Materials Sciences and Engineering Division, Basic Energy Sciences, Office of Science, US Dept.~of Energy.  The author acknowledges several discussions with John Mitchell about the topics addressed in this paper.
\end{acknowledgments}

\bibliography{references}

\end{document}